\documentclass[
 aip,
 amsmath,amssymb,
 reprint,
]{revtex4-2}
\usepackage{graphicx}

\usepackage{dcolumn}
\usepackage{bm}

\usepackage[utf8]{inputenc}
\usepackage[T1]{fontenc}
\usepackage{mathptmx}
\usepackage{etoolbox}
\usepackage{subcaption}
\usepackage[colorlinks,citecolor=blue,urlcolor=blue,linkcolor=blue]{hyperref}

\newcommand{\HR}[1]{{\color{green} #1}}

\begin{document}

\title{Dilatancy-induced surface deformation in dense cohesive granular media}
\author{Huzaif Rahim}
\author{Thorsten P\"oschel}
\author{Sudeshna Roy}
\email{sudeshna.roy@fau.de}
\affiliation{
Institute for Multiscale Simulation, \\ Friedrich-Alexander-Universit\"at Erlangen-N\"urnberg, \\ Cauerstrasse 3, 91058 Erlangen, Germany
}
\date{\today}

\begin{abstract}
When granular materials with interstitial liquid bridges are sheared in a split-bottom cell, a localized shear band develops, accompanied by a surface elevation. Cohesion, governed by the surface tension of the interstitial liquid, enhances dilatancy in dense cohesive packings, leading to expansion within the shear band and the emergence of a surface elevation. Surface deformation is observed not only in cohesive systems with high particle density and large liquid surface tension, but also in those with lower values of these parameters. The equivalent Bond number arises as a key control parameter for the surface deformation, shaping both the evolution of the surface profile and the packing density. At higher shear rates, inertial effects dominate dilatancy, resulting in less pronounced surface deformation.
\end{abstract}

\maketitle

\section{Introduction}

Cohesive granular materials have attracted considerable attention owing to their significance in industrial applications and geophysical phenomena.\cite{cunez2024particle, torres2017flowability} Particles in cohesive granular materials experience both cohesive and frictional interactions, with cohesion amplifying the frictional contribution to dilation. Cohesion, therefore, plays a critical role in determining the response of these materials to external forcing under shear.\cite{roy2017effect, shi2020steady, fournier2005mechanical, khamseh2015flow}

Cohesive particles typically pack less densely than non-cohesive\cite{roy2017effect}\HR{.} For example, the packing density of dry glass beads is around 0.6 but decreases to $\sim 0.3$ when cohesion is present.\cite{enferad2021powder, wang2021packing} Strong cohesion disrupts uniform packing, leading to particle clustering and non-uniform density, which in turn causes shear band widening, velocity fluctuations, and dilation or compression.\cite{askarishahi2022capability, roy2017effect, gans2020cohesion}

A notable phenomenon that has attracted considerable attention is the emergence of secondary flows driven by dilation in granular shear flows. Previous studies by Dsouza and Nott\cite{dsouza2021dilatancy} and Krishnaraj and Nott\cite{krishnaraj2016dilation} provided key insights into the secondary flows of frictional particles in split-bottom shear cells. Building on this foundation, the present work examines flows in cohesive granular materials, with particular emphasis on the role of cohesion-induced dilation.

We investigate the interplay between cohesion and dilation in wet granular materials. Cohesion is introduced through capillary bridges, with the liquid’s surface tension varied to control its intensity. By systematically varying cohesion within the granular assembly, we analyze its influence on dilation and flow patterns. We also examine how shear rate affects flow behavior and dilation in cohesive granular materials.

\section{Model}

\subsection{Split-bottom shear cell}

\begin{figure}[htb]
  \begin{center}
\includegraphics[width=\columnwidth]{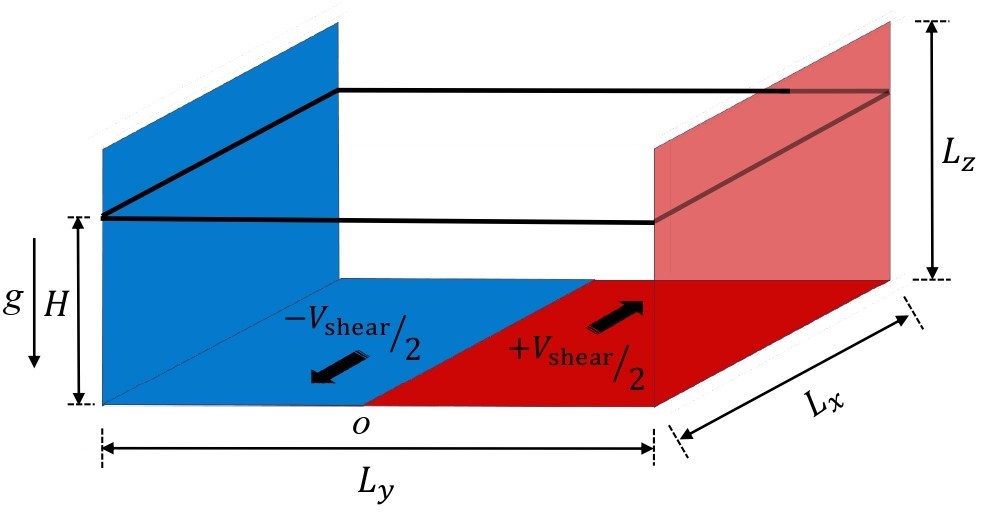}
  \end{center}
\caption{Linear split-bottom shear cell}
\label{fig:Geometry}
 \end{figure}

The linear split-bottom shear cell (LSC) consists of two L-shaped walls that slide past each other at velocities $\pm V_\text{shear}/2$. The system size is $(L_{x}, L_{y}, L_{z}) = (20, 80, 25)d_\text{p}$, with filling height $H \approx 20d_\text{p}$, where $d_\text{p}$ is the mean particle diameter. Gravity $g$ acts in the negative $z$-direction. \autoref{fig:Geometry} sketches the shear cell and defines the coordinate system. This is a well-known geometrical setup for studying {the} rheology of granular materials; similar geometry is used in many previous works.\cite{rahim2024alignment, ries2007shear, depken2007stresses}

The particle diameters are uniformly distributed in $[0.83,1.24]\,\text{mm}$ with mean $d_\text{p}=1.035\,\text{mm}$. 
The total volume of particles inserted into the simulation domain is scaled by a packing efficiency factor of $0.64$, which represents the expected solid volume fraction after random deposition and avoids unrealistically dense initial packings. 

\subsection{Particle contact model}
\label{app:non-linear contact model}
\subsubsection{Hertz‑Mindlin visco‑elastic contact model}
For simulating particle dynamics, we use the discrete element (DEM) program MercuryDPM.\cite{weinhart2020fast} For the details of DEM, we refer, e.g., to Refs. \onlinecite{matuttis2014understanding, poschel2005computational}. For
the particle-particle interaction force, we assume the Hertz-Mindlin no-slip contact model.\cite{di2004comparison} The normal force component is\cite{brilliantov1996model,poschel2005computational}
\begin{equation}
    \vec{F}^\text{n}_{ij} = \min\left(0, -\rho \xi_{i,j}^{3/2} - \frac{3}{2}A^\text{n}_{i,j}\rho\sqrt{\xi_{i,j}}\dot{\xi}_{i,j}\right) \vec{e}^\text{\,n}\,,
\end{equation}
where $\xi_{i,j} = R_i + R_j - \vert\vec{r}_i-\vec{r}_j\vert$ is the compression of two interacting particles $i$, $j$ of radii $R_i$ and $R_j$ at positions $\vec{r}_i$ and $\vec{r}_j$ and $\vec{e}^\text{\,n} = (\vec{r}_i-\vec{r}_j)/\vert\vec{r}_i-\vec{r}_j\vert$ is the normal unit vector. Further, we use the damping
constant $A^\text{n}_{i,j} = 5\times 10^{-6}$ s. This value corresponds to a coefficient of restitution of $0.7$ for two granular particles with
elastic modulus $E = 5$ MPa and diameter $d_\text{p}$ impacting
at a velocity of $0.016\,\text{m/s}$.\cite{muller2011collision}For higher impact velocities (up to 
$0.16$ m/s), the damping constant is recalculated accordingly to maintain a consistent coefficient of restitution.

The elastic constant is given by \cite{brilliantov1996model}
\begin{equation}
    \rho_{i,j} = \frac{4}{3} \, E^*_{i,j} \, \sqrt{R^*} 
    \label{eq:Hertz_rho}
\end{equation}
where $E^*_{i,j}$ is the effective elastic modulus, and $R^*$ is the effective radius. 

We model the tangential viscoelastic forces using a path-dependent approach, which accounts for the history of tangential displacement between contacting particles. This follows the no-slip solution of Mindlin \cite{mindlin1949compliance} for the elastic component and the formulation by Parteli and Pöschel \cite{parteli2016particle} for the tangential dissipative constant ($A^\text{t}_{i,j} \approx 2 A^\text{n}_{i,j}\, E^*_{i,j}$). The tangential force is capped by the static friction limit between particles (Coulomb law),

\begin{equation}
    \vec{F}^\text{t}_{i,j} = -\min \!\!\!\left[\! \mu\left|\vec{F}^\text{n}_{i,j}\right|, \!\! \int\limits_\text{path}^{}\!\!\! 8 G^*_{i,j}\sqrt{R^*_{i,j} \xi_{i,j}} \,\text{d}s 
    + A^\text{t}_{i,j}  \sqrt{R^*_{i,j} \xi_{i,j}} v^\text{t} \!\right]\!\! \vec{e}^\text{\,t}\,,
\end{equation}
where $\mu$ is the friction coefficient, $G^*_{i,j}$ is the effective shear modulus. 
The integral is evaluated over the tangential displacement path, where
$ds = |\mathbf{v}_{i,j}^t|\,dt$ is the incremental tangential relative displacement at the contact point, accumulated from the onset of contact until separation.\cite{parteli2016particle}

\subsubsection{Liquid bridge contact model}
The nonlinear capillary force depends on particle size, contact properties, liquid properties, and the liquid saturation level in the system. Specifically, it is governed by three key parameters: the surface tension, $\gamma$, which determines the maximum adhesive force, the contact angle, $\theta$, and the liquid bridge volume, $V_\text{b}$, which also defines the maximum interaction distance between particles at the point of bridge rupture (rupture distance).\cite{willett2000capillary, bagheri2024approximate, bagheri2025discrete} The adhesive capillary force between particles $i$ and $j$, denoted $\vec{F}_{i,j}^\text{c}$,  is modeled using the approximations by Willett et al. \cite{willett2000capillary} based on the particle specifications, contact properties, liquid properties, and liquid saturation in the system,
\begin{equation}
\vec{F}_{i,j}^\text{c} = \frac{F^\text{c}_\text{max}}{\sqrt{1+1.05\overline{S}_{i,j}+2.5\overline{S}_{i,j}^2}}\vec{e}^\text{\,n}\,,
\end{equation}
with the normalized distance {between} the surfaces of the particles 

\begin{equation}
\overline{S}_{i,j} = \max\left(0,-\xi\right)\sqrt{\frac{d_\text{p}}{V_\text{b}}}
\end{equation}
and $F^\text{c}_\text{max}=\pi{d_p}\gamma\cos{\theta}$. To study the effect of cohesion in unsaturated granular materials, we simulate the system for a range of surface tensions, $\gamma \in \{0, 0.02, 0.04, 0.06, 0.08, 0.16\}\,\text{N/m}$. 

\subsection{Material and process parameters}
Unless stated otherwise, all simulations were performed using the standard parameters listed in \autoref{tab:material_parameters}.
In selected simulations\HR{,} alternative values of the shear rate, particle density, or surface tension were employed.
\begin{table}[h!]
\caption{DEM standard simulation parameters}
\label{tab:material_parameters}
\begin{center}
\begin{tabular}{l@{\quad}l@{\quad}l}
\hline
variable & unit & value\\ \hline
                   elastic modulus ($E$)  & MPa& $5$\\
                   sliding friction coeff. ($\mu_s$)  & -& $0.10$\\
                   rolling friction coeff. ($\mu_r$)  & -& $0.005$\\
                   contact angle ($\theta$)  & degree& $20$\\
                   liquid bridge volume ($V_\text{b}$)  & nl& $100$\\                   
                   particle density ($\rho_\text{p}$)  & $\text{kg/m}^3$& $2000$\\
                   surface tension ($\gamma$)  & $\text{N/m}$ & $0-0.16$\\
shear rate ($V_\text{shear}$) & $\text{m/s}$ & $0.016$\\
\hline
\end{tabular}
\end{center}
\end{table}

\section{Coarse graining}
We employ the coarse-graining scheme by Strobl et al.\cite{strobl2016exact} as implemented in MercuryCG \cite{weinhart2016influence} to compute the macroscopic fields through the exact evaluation of the intersections between the spheres and the triangulated domain. This approach yields a precise value of the local solid fraction, enabling highly accurate measurements of the packing density. We simulated the process for $300\,\text{s}$. The macroscopic fields are then obtained by averaging over the (periodic) $x$-direction during the interval $t\in (250, 300)\,\text{s}$, after the system has reached its stationary state.

\section{Surface deformation at slow shear}
\subsection{Packing density characteristics}
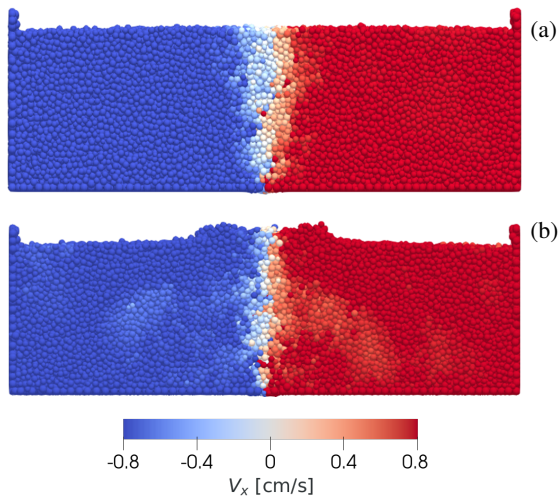
\begin{figure}[b]
   \centering
    \begin{subfigure}{\columnwidth}
     \includegraphics[width=0.8\columnwidth]{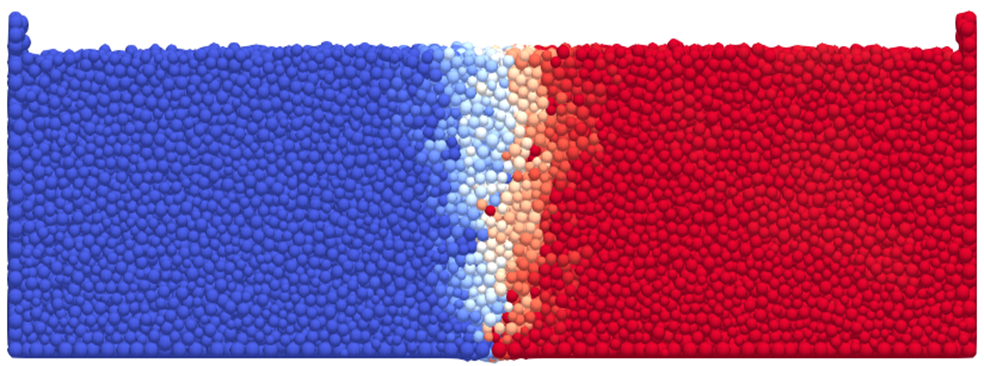}       \begin{picture}(0,0)
            \put(0,60){\small{(a)}}
        \end{picture}
    \end{subfigure}
    \begin{subfigure}{\columnwidth}

    \includegraphics[width=0.8\columnwidth]{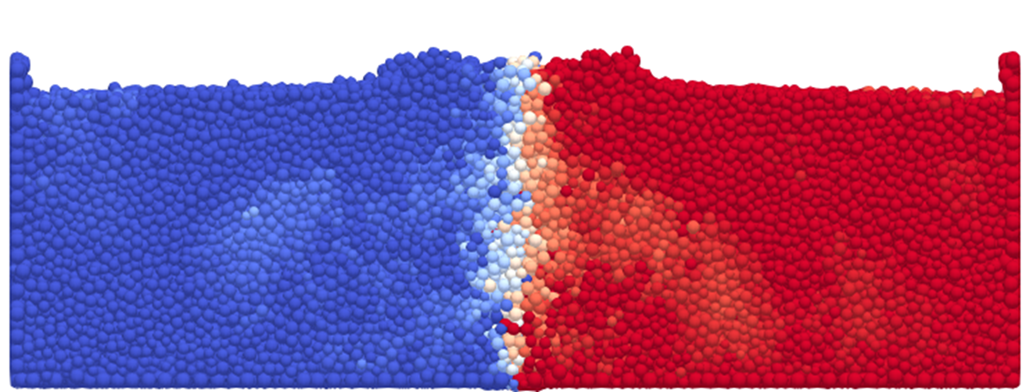} 
        \begin{picture}(0,0)
            \put(0,60){\small{(b)}}
        \end{picture}
    \end{subfigure}
    \begin{subfigure}{\columnwidth}
        \centering
        \includegraphics[width=0.5\columnwidth,bb=0 10 470 150,clip]{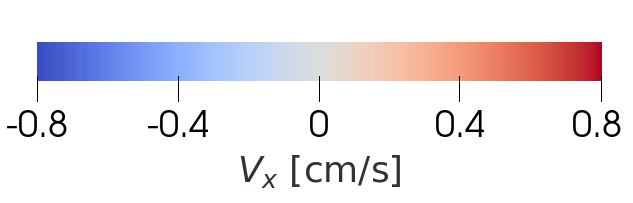}
    \end{subfigure}
  
    \caption{Snapshots of DEM simulations of a linear shear cell ($y-z$ plane) in the stationary state for (a) cohesionless materials with $\text{Bo}_g = 0$ and (b) highly cohesive material with $\text{Bo}_g = 6.35$. The color codes for the velocity component in {the} shear direction.}
    \label{fig:ObservationHeap} 
\end{figure}

\autoref{fig:ObservationHeap} shows snapshots of DEM simulations of a linear shear cell for (a) cohesionless material with surface tension $\gamma = 0$ and (b) highly cohesive material with $\gamma = 0.16\,\text{N/m}$. Color codes for the velocity {component} in shear direction, $v_x$.
For cohesive material, we observe a pronounced accumulation of particles above the shear band, while for non-cohesive material\HR{,} the surface remains flat. This behavior is characteristic of shear of cohesive granular matter.

\begin{figure}[!htb]
 \includegraphics[width=\columnwidth,
        trim=0cm 0cm 0cm 0cm,
        clip]{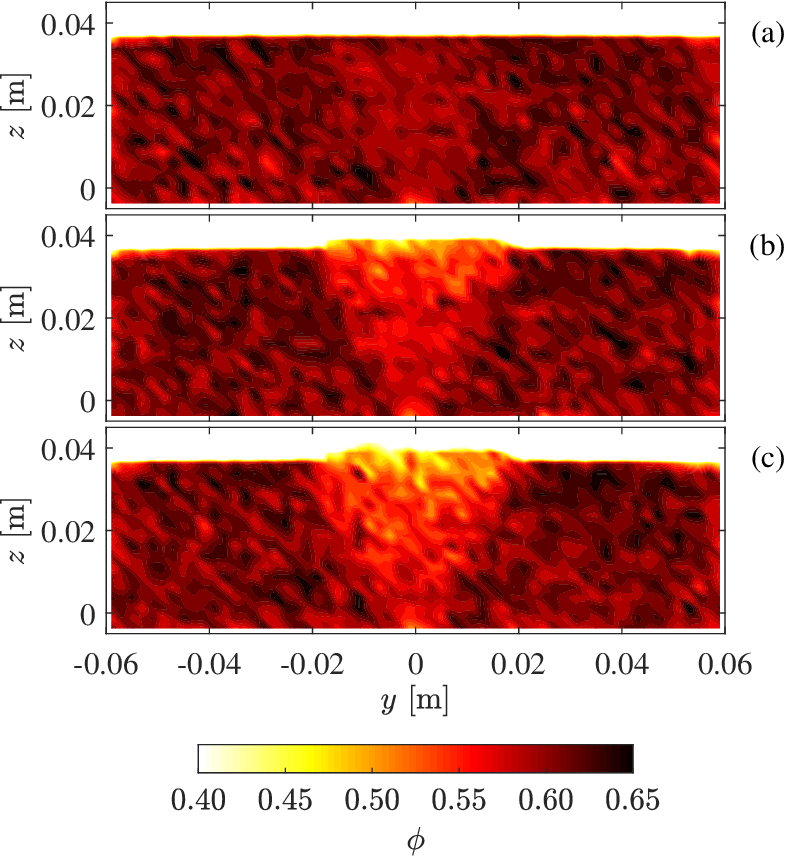}
    \caption{Field of the packing density $\phi$ for (a) $\gamma = 0$ (b) $\gamma = 0.08\,\text{N/m}$ and (c) $\gamma = 0.16\,\text{N/m}$, particle density $\rho_\text{p} = 2000\,\text{kg/m}^3$ and shear velocity $V_\text{shear} = 0.016\,\text{m/s}$.} 
    \label{fig:Dilatancy} 
\end{figure}

Interparticle cohesion changes the microstructure due to rolling and sliding friction.\cite{gilabert2007computer, fournier2005mechanical, roy2017effect} \autoref{fig:Dilatancy} shows the field of packing density in the regime of stationary velocity, $\phi(y, z)$, averaged over interval $t\in (250, 300)\,\text{s}$, for cohesionless, moderately and highly cohesive material, $\gamma=\{0, 0.08, 0.16\}$.
For cohesive material\HR{,} a shear band forms, indicated by a V-shaped region of low density and corresponding surface deformation (\autoref{fig:Dilatancy}(b,c)). 

\subsection{2D projected area of surface profile}

Although the free surface exhibits spatial fluctuations along the shear direction $x$, the projected cross-sectional area 
$A(t)$ provides a scalar measure of surface deformation intensity. For the present system, variations in surface roughness along $x$ affect the absolute surface area but do not alter the monotonic relationship between the projected area and the degree of shear-induced surface deformation.

For calculating the projected area, we discretize the domain into $50$ equal-sized strips in the shear direction of width $\Delta y$. For each bin centered at $y_j$, the maximum particle height is determined
locally along the shear direction $x$ and subsequently averaged over $x$, yielding the projected surface profile
\begin{equation}
z_{\text{surf}}(y_j,t)
=
\left\langle
z_{p,\text{max}}(x,y_j,t)
\right\rangle_x ,
\label{eq:z-surf}
\end{equation}
where $z_{p,\text{max}}(x,y_j,t)$ is the maximum vertical coordinate of particles at a given $x$ within the bin. This yields a mean cross-sectional envelope of the free surface in the $yz$-plane.  Since variations of the surface along the shear direction are not resolved, $z_{\text{surf}}(y,t)$ does not represent the true three-dimensional surface area but a projected, cross-sectional measure that constitutes a lower bound. 

The 2D projected area is then obtained by numerical integration,
\begin{equation}
A(t)
=
\int_0^{L_y} z_{\text{surf}}(y,t)\,\mathrm{d}y ,
\label{eq:A-surf}
\end{equation}

{which is evaluated using Simpson’s rule at the bin centers $y_j$.
For the subsequent analysis, $A(t)$ is normalized by
$A_0 \equiv A(t=0)$, the initial value prior to shear-induced
deformation.

}

\begin{figure}[!htb]
\includegraphics[width=\columnwidth]{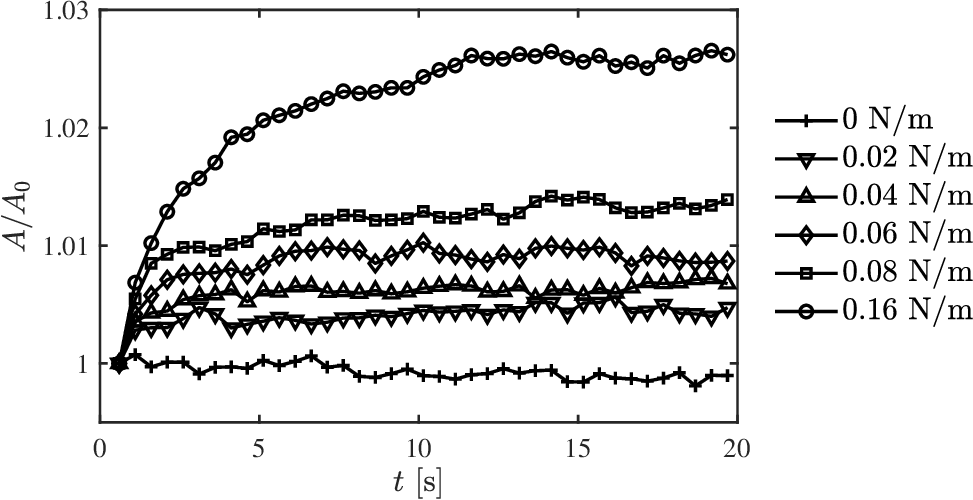}
\caption{Evolution of the surface area $A/A_0$ for the shear rate $V_\text{shear} = 0.016\,\text{m/s}$. The values of the liquid surface tension, $\gamma$, are given in the legend.}
\label{fig:heapEvolution}
\end{figure}

\autoref{fig:heapEvolution} shows the surface area ${A}/{A_{0}}$ as a function of time for different surface tension{s} of the liquid. We see that after about $10\,\text{s}$ the surface area has assumed its stationary value. While for non-cohesive material $A/A_0$ essentially remanins at its initial value, for cohesive material under shear, the surface expands due to developed dilatancy.

\subsection{Relevance of the Bond number}
\label{sec:Bond}

\begin{figure}[b]
   \centering
    \begin{subfigure}{\columnwidth}
    \includegraphics[width=0.8\columnwidth]{Figures/ST5Image.png} 
                \begin{picture}(0,0)
            \put(0,60){\small{(a)}}
        \end{picture}
    \end{subfigure}\hfill
    \vspace{2mm}
    \begin{subfigure}{\columnwidth}
    \includegraphics[width=0.8\columnwidth]{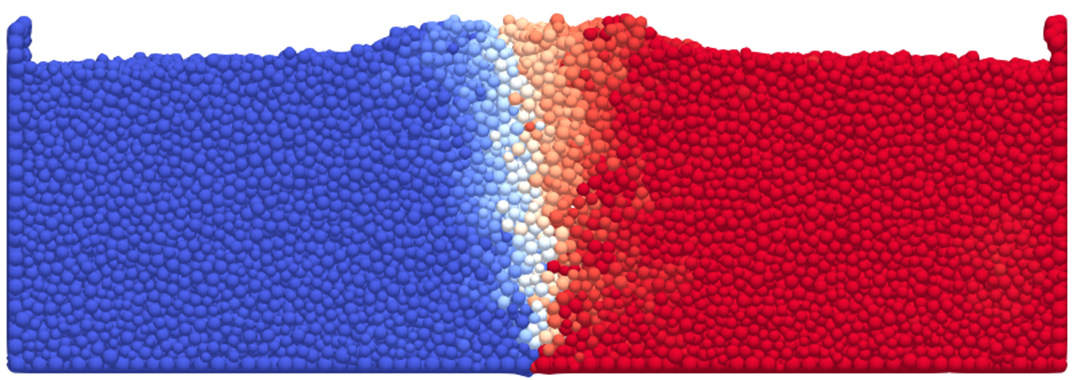}     
        \begin{picture}(0,0)
            \put(0,60){\small{(b)}}
        \end{picture}
    \end{subfigure}
    \begin{subfigure}{\columnwidth}
        \centering
        \includegraphics[width=0.5\columnwidth,bb=0 10 470 150,clip]{Figures/ColorBar.png}
    \end{subfigure}
    \caption{Snapshots of simulations of steady-state granular shear flow at the same Bond number, $\text{Bo}_g = 6.35$. (a) $\rho_\text{p} = 2000\,\text{kg/m}^3$ and $\gamma = 0.16\,\text{N/m}$ and (b) $\rho_\text{p} = 850\,\text{kg/m}^3$ and $\gamma = 0.07\,\text{N/m}$.}
    \label{fig:ObservationHeapBo} 
\end{figure}

\begin{figure}[!htb]
\includegraphics[width=\columnwidth,
        trim=0cm 0cm 0cm 0cm,
        clip]{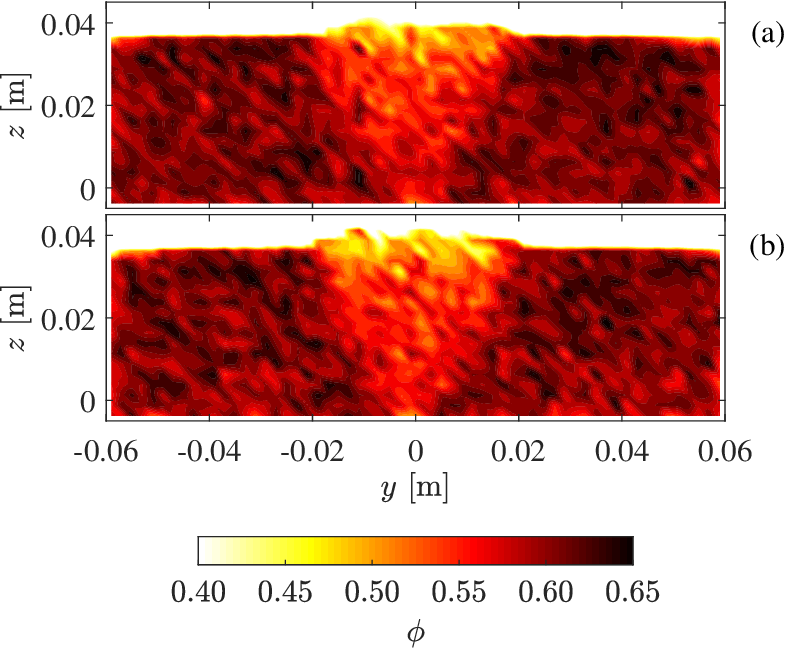}
    \caption{Field of the packing density corresponding to \autoref{fig:ObservationHeapBo}.} 
    \label{fig:LocalDensityContourBo}
\end{figure}

\begin{figure}[!htb]
   \includegraphics[width=0.8\columnwidth]{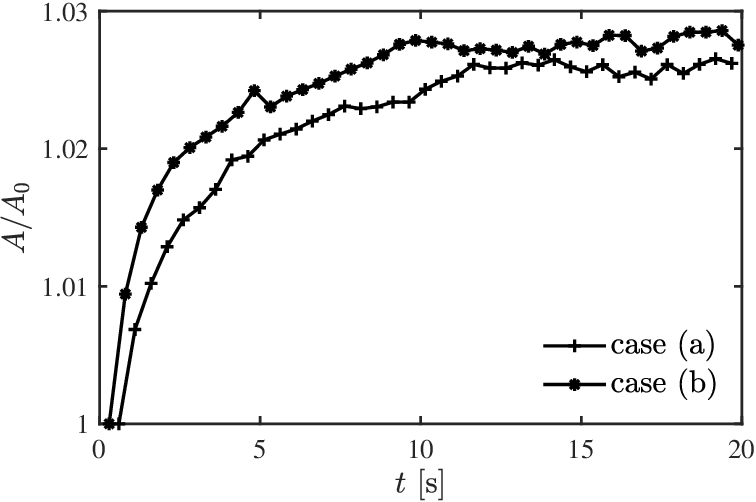}
    \caption{Evolution of exposed surface area $A/A_0$ corresponding to \autoref{fig:ObservationHeapBo}.}
    \label{fig:HeapEvolutionBo} 
\end{figure}

The Bond number 
\begin{equation}
    \text{Bo}\equiv \frac{9\gamma}{4\rho_\text{p}d_\text{p}^2g}
\end{equation}
quantifies the effect of cohesion compared to volume forces (here\HR{,} gravity). To study the influence of cohesion on {granular} shear flow, we performed simulations with varying parameters, where we kept the Bond number invariant. \autoref{fig:ObservationHeapBo}
shows snapshots of simulations of steady-state flow for two cases (a) $\rho_\text{p} = 2000\,\text{kg/m}^3$ and $\gamma = 0.16\,\text{N/m}$ (case (a)) and (b) $\rho_\text{p} = 850\,\text{kg/m}^3$ and $\gamma = 0.07\,\text{N/m}$ (case (b)), both corresponding to $\text{Bo}_g = 6.35$. In both cases\HR{,} we see similar deformation of the granular surface, underscoring the relevance of the Bond number for the system's macroscopic behavior. The same conclusion can be drawn from the field of packing density shown in \autoref{fig:LocalDensityContourBo}, corresponding to \autoref{fig:ObservationHeapBo}. Despite different simulation parameters (but {the} same Bond number), the density fields remain approximately invariant.
The evolution of the cross-sectional area, in \autoref{fig:HeapEvolutionBo}, obtained from \autoref{eq:A-surf} also shows a similar trend for the two cases investigated.

From all these observations, we conclude that the characteristics of {the} shear flow of adhesive granular material are governed by the value of the Bond number. Other combinations of particle material density and fluid surface tension (not shown here) leading to the same Bond number show similar behavior.

\section{Surface deformation at rapid shear}

\begin{figure}[!htb]
\includegraphics[width=\columnwidth,
        trim=0cm 0cm 0cm 0cm,
        clip]{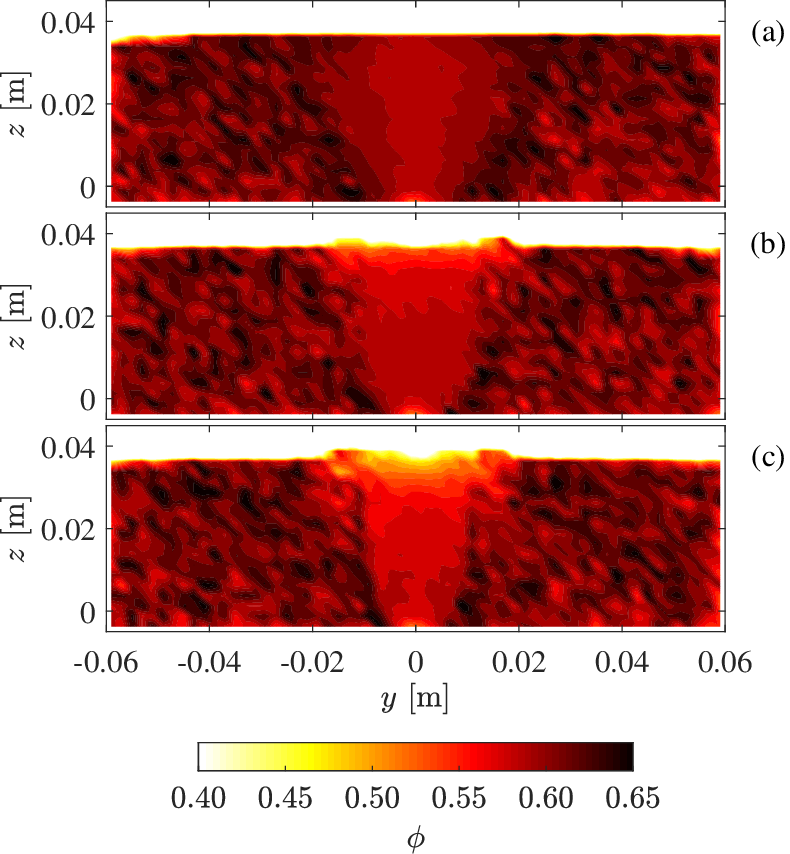}
    \caption{Field of the packing density for different fluid surface tension. (a) $\gamma = 0$, (b) $\gamma = 0.08\,\text{N/m}$, (c) $\gamma = 0.16\,\text{N/m}$, 
    and shear velocity $V_\text{shear} = 0.16\,\text{m/s}$. }
    \label{fig:FSDilatancy} 
\end{figure}

\begin{figure}[htb]
        \includegraphics[width=\columnwidth]{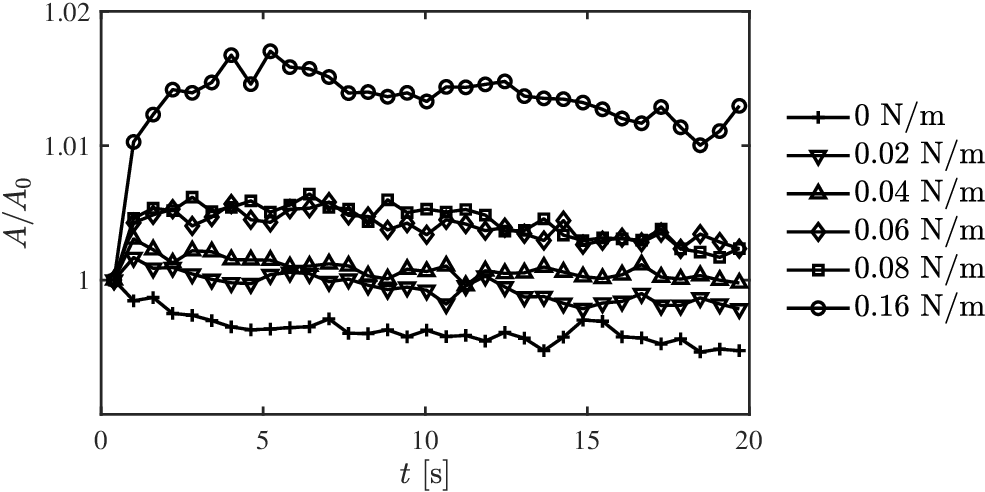}
  
    \caption{Exposed granular surface, $A/A_0$, as a function of time for rapid shear at $V_\text{shear} = 0.16\,\text{m/s}$ and different values of the liquid's surface tension as given in the legend.} 
    \label{fig:heapEvolutionFastShear} 
\end{figure}

For cohesionless materials, $\gamma = 0$, at $V_\text{shear} = 0.16\,\text{m/s}$, the field of packing density, \autoref{fig:FSDilatancy}(a), reveals much more pronounced dilatiation in the shear band region, compared to low shear velocity, $V_\text{shear} = 0.016\,\text{m/s}$, shown in \autoref{fig:Dilatancy}(a). Comparing \autoref{fig:FSDilatancy}(a-c), we see that increased surface tension leads to more intense dilation in the shear band region. This effect is, however, less pronounced than for slow shear, $V_\text{shear} = 0.016\,\text{m/s}$, shown in \autoref{fig:Dilatancy}(a-c). The deformation of the granular surface due to shear is less prominent compared to slow shear, see \autoref{fig:heapEvolutionFastShear} 
showing the exposed granular surface area, $A/A_0$, as a function of time for $V_\text{shear} = 0.16\,\text{m/s}$. For comparison, \autoref{fig:heapEvolution} shows the corresponding data for slow shear at $V_\text{shear} = 0.016\,\text{m/s}$.

\section{Shear dilatancy and Bond number}

\begin{figure}[htb]
  \begin{subfigure}{\columnwidth}
     \includegraphics[width=0.8\columnwidth]{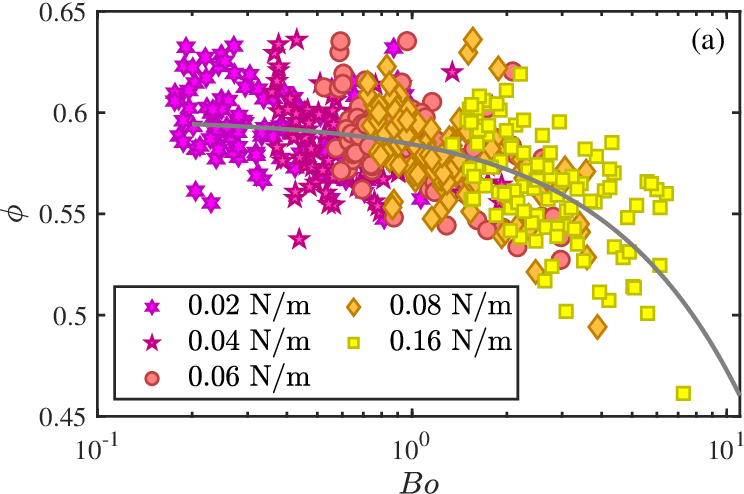}
  \end{subfigure}\hfill    
  \vspace{2mm}
  \begin{subfigure}{\columnwidth}
      \includegraphics[width=0.8\columnwidth]{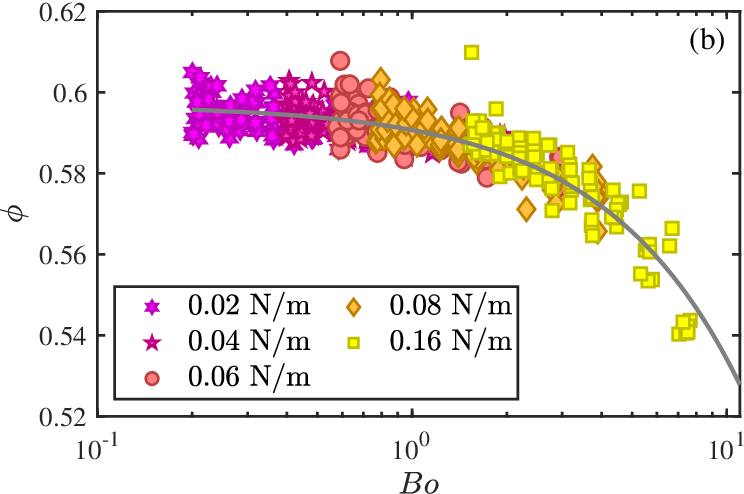}
  \end{subfigure}    
  \caption{Packing fraction as a function of the local Bond number for different cohesion and shear velocity $V_\text{shear} = 0.016\,\text{m/s}$ (a)  and $V_\text{shear} = 0.16\,\text{m/s}$ (b). The values of the liquid surface tension, $\gamma$, are given in the legend. The gray lines show \autoref{LinearFit} with $\text{Bo}_\text{c} = 80.58$ (a) and $\text{Bo}_\text{c} = 159.11$ (b).}
 \label{fig:cohesiondilatancy} 
\end{figure}

We consider the field of particle packing fraction, $\phi$, in the region of pronounced shear (shear band center). This region is defined by the coordinates in the $y-z$ plane where the local $z$-dependend strain rate exceeds 80\% of the corresponding maximal value, $\dot{\epsilon}_\text{max}(z)$:\cite{roy2017general}
\begin{equation}
    \dot{\epsilon}(z) \ge 0.8\, \dot{\epsilon}_\text{max}(z)\,.
\end{equation}
\autoref{fig:cohesiondilatancy} 
shows the packing fraction as a function of the Bond number. The simulation parameters cover the entire range of cohesion considered in this paper. We find that the data are consistent with a generic function depending solely on \text{Bo}, 
\begin{equation}\label{LinearFit}
\phi = \phi_0 \left(1 - \frac{\text{Bo}}{\text{Bo}_\text{c}}\right)\,,
\end{equation}
where $\phi_0 \approx 0.60$ corresponds to the packing density of non-cohesive granular flows, and $\text{Bo}_\text{c}$ is a fit parameter. Larger values of $\text{Bo}_\text{c}$ imply stronger resistance to dilation. These results indicate that for slow shear velocity, cohesive forces dominate{s} while at faster shear velocity, inertial effects dominate cohesion, reducing dilatation and granular surface deformation.

\section{Conclusions}

We studied the behaviour of dense cohesive granular matter under linear shear, in a split-bottom shear cell with periodic boundary conditions in {the} shear direction. Cohesion was modeled by liquid bridges spanning between particles in contact or {in} close vicinity. The degree of adhesion was controlled by the coefficient of surface tension of the liquid. We observe the formation of a low-intensity shear zone and corresponding deformation of the free granular surface according to the granulate's expansion. Slow shear results in pronounced surface deformation, whereas at higher shear rates the influence of cohesion is increasingly dominated by inertial effects, leading to reduced surface deformation.

For both cases, slow and rapid shear, the data collapse for simulation at different adhesion (liquid's surfqce tension) and different material density, when the drawing the characteristics as functions of the Bond number, such as the free granular surface area $A/A_0 (\text{Bo})$ and packing fraction in the shear-band region, $\phi(\text{Bo})$.

\section*{Acknowledgement}
We thank Holger G\"otz and Hongyi Xiao for stimulating discussion. This work was supported by the German Science Foundation (DFG) through grant PO472/40-1. 

\bibliography{aipsamp}
\end{document}